\documentclass[aps,prd,preprint,superscriptaddress]{revtex4-1}
\usepackage[mathscr]{euscript}
\usepackage{float,epsfig}
\usepackage{bm}
\usepackage{graphicx}
\usepackage{subfigure}
\usepackage{amsmath,amssymb,amsthm}
\usepackage[colorlinks=true,linkcolor=blue,urlcolor=blue]{hyperref}
\newcommand{\bea}{\begin{eqnarray}}
\newcommand{\eea}{\end{eqnarray}}
\newcommand{\beq}{\begin{equation}}
\newcommand{\eeq}{\end{equation}}

\def\/{\over}

\begin{document}


\title{Dynamical gravitational Casimir-Polder interaction}


\author{Yongshun Hu}
\email[Corresponding author. ]{huys@cqupt.edu.cn}
\affiliation{School of Electronic Science and Engineering, Chongqing University of Posts and Telecommunications, Chongqing 400065, China}
\author{Dan Wen}
\email[Corresponding author. ]{wendan@cqupt.edu.cn}
\affiliation{School of Electronic Science and Engineering, Chongqing University of Posts and Telecommunications, Chongqing 400065, China}




\begin{abstract}
We explore the time-dependent Casimir-Polder-like quantum gravitational interaction between a nonpointlike object and a gravitational Dirichlet boundary, i.e., the dynamical gravitational Casimir-Polder interaction, based on the theory of linearized quantum gravity.
We demonstrate that the dynamical interaction potential is nonzero prior to the radiation, which is generated by the gravitational vacuum-fluctuation-induced mass quadrupole of the object, being reflected by the gravitational boundary and back-reacted to the object (i.e., the mirror image of the object lies outside its causal region). This indicates the nonlocality of the fluctuating gravitational field in vacuum and calls for a reevaluation of the inherent causality within the interaction.
Moreover, the dynamical gravitational Casimir-Polder interaction can be either attractive or repulsive depending on the distance of the object with respect to the boundary and the time of interaction.
When the interaction time is sufficiently long for the system to approach asymptotic equilibrium, the dynamical gravitational Casimir-Polder interaction potential reduces to the static one, which is time-independent and consistently repulsive in both the near and far regimes.

\end{abstract}


\maketitle

\section{Introduction}
\label{sec_in}
\setcounter{equation}{0}
Quantum vacuum fluctuations are an inevitable consequence of quantum theory and may induce some observable effects which are not predicted by classical physics. A well-known example is the static electromagnetic Casimir-Polder (CP) interaction, which arises from the dipolar interaction between a neutral atom and a perfectly conducting surface \cite{CP,power1982}. Physically speaking, electromagnetic vacuum fluctuations induce instantaneous electric dipole moment in neutral atom, which subsequently radiates an electromagnetic field and couples with the fields reflected by the conducting plane boundary, and an interaction is thus induced. Intuitively, the CP interaction can be regarded as the interaction between an atom and its mirror-image counterpart, where the conducting surface acts as the mirror. In recent years, such CP effects have been extensively investigated under diverse conditions and applied in various areas (see Refs.  \cite{Klimchitskaya2009,Woods2016} for recent reviews). Also, it is worth noting that the measurement of the CP interaction has already been accomplished with precision \cite{Sukenik1993,Druzhinina2003,Bender2010}.

Likewise, if gravity is quantized, a CP-like quantum gravitational interaction may also exist between a nonpointlike object and a gravitational boundary. Unfortunately, a comprehensive theory of quantum gravity remains elusive at present. Nevertheless, one can still study low-energy quantum gravitational effects since, at low energy scales, general relativity can be treated as an effective field theory or linearly quantized. For example, by applying the framework of effective field theory, the quantum correction to the Newtonian gravitational potential between two point masses has been derived
\cite{Donoghue1994prl,Donoghue1994prd,Hamber1995,Kirilin2002,Holstein2003,Holstein2005}. Within the context of linearized quantum gravity, the static quantum gravitational CP interaction between a nonpointlike object and a gravitational plane boundary has been obtained \cite{Hu2017}, which behaves as $z^{-5}$ and $z^{-6}$ in the near and far regimes respectively with $z$ being the distance between the object and the boundary. Here the static interaction refers to a condition where the interacting system is in equilibrium and is characterized by a time-independent potential.
Note here that, although ordinary materials can hardly reflect gravitational waves \cite{Smolin1985}, quantum matter such as superconducting films might enable such reflection \cite{Chiao2017,Minter2010}.

Recently, a dynamical (time-dependent) formalism for the CP-like quantum gravitational interaction potential between two entangled nonpointlike objects in the presence of gravitational boundaries is presented in Ref. \cite{yongs2023}. This formalism demonstrates that the interobject interaction potential is nonzero only when one object (or its image) lies within the light cone of the other.
Consequently, a natural question arises as to whether the dynamical gravitational CP interaction potential between a nonpointlike object and a gravitational boundary appears exclusively when the mirror image of the object lies within its light cone.
In this paper, we explore the time-dependent quantum gravitational interaction between a nonpointlike object and a gravitational Dirichlet boundary, i.e., the dynamical gravitational CP interaction, based on the theory of linearized quantum gravity. First, we give a brief description of the formulae for the interaction under the method proposed by Dalibard, Dupont-Roc, and Cohen-Tannoudji (DDC)~\cite{DDC1982,DDC1984}. Then, we derive the general formalism of the dynamical gravitational CP interaction potential and analyze the resulting expressions under specific conditions.
Throughout this paper, the Greek indices run from $0$ to $3$, the Latin indices run from $1$ to $3$, and the Einstein summation convention for repeated indices is assumed.

\section{Basic equations}
\label{sec_ba}
Consider a ground-state nonpointlike object A interacting with the fluctuating gravitational fields in vacuum near a gravitational plane boundary. Generally, let us model the object A as a multilevel system where the ground and excited states are denoted as $|e_0\rangle$ and $|e_s\rangle ~( s=1,2,3,...)$ respectively, with the corresponding energy spacing labeled as $\hbar \omega_s$. The total Hamiltonian of the system is
\beq
H=H_S+H_F+H_I\;,
\eeq
where $H_{S}$ is the Hamiltonian of the object A, $H_{F}$ is the Hamiltonian of the gravitational fields in vacuum, and $H_I$ the interaction Hamiltonian between the object and the gravitational fields. Here $H_I$ takes the form
\beq
H_I=-\frac{1}{2}Q_{ij}E_{ij}(\vec r_A)\;,
\eeq
where $Q_{ij}$ is the quadrupole moment operator of object A,  and $E_{ij}$ is the gravitoelectric tensor of the fluctuating gravitational fields in vacuum. By drawing an analogy between the linearized Einstein field equations and the Maxwell equations, the gravitoelectric tensor can be defined as $E_{ij}=-c^2 C_{0i0j}$ \cite{Campbell1971,Campbell1976,Maartens1998,Matte1953,Ramos2010,Szekeres1971,Ruggiero2002}, where $C_{\mu\nu\alpha\beta}$ represents the Weyl tensor and $c$ denotes the speed of light. Under the weak-field approximation, we expand the metric tensor for the fluctuating gravitational fields as a sum of the flat spacetime metric $\eta_{\mu\nu}$ and a linearized perturbation $h_{\mu\nu}$. Then, $E_{ij}$ can be expressed as
\beq\label{Eij}
E_{ij}=\frac{1}{2}\ddot h_{ij}\;.
\eeq
where a dot denotes the derivative with respect to time $t$. In the presence of a gravitational Dirichlet boundary, the quantized metric perturbations can be expressed as (in the transverse traceless gauge)~\cite{yu2018}
\beq\label{hij}
h_{ij}=\sum_{\vec p, \lambda}h^{\lambda}_{ij,\vec p}=\sum_{\vec p, \lambda}\sqrt{\frac{\hbar G}{2\pi^2c^2\omega}}\left\{ a_{\lambda}(\omega)\left[e^{(\lambda)}_{ij}(\vec p) e^{i(\vec{p}\cdot\vec{x}-\omega t)}-e^{(\lambda)}_{ij}(\vec{p}_{-}) e^{i(\vec{p}_{-}\cdot\vec{x}-\omega t)}\right]+\text{H.c.} \right\}\;,
\eeq
where the gravitational plane boundary is assumed to be placed at $z=0$, and the metric perturbation satisfies $h^{\lambda}_{ij,\vec p}|_{z=0}=0$.  Here $\hbar$ is the reduced Planck constant, $G$ is the Newtonian gravitational constant, $a_{\lambda}(\omega)$ represents the annihilation operator of the fluctuating gravitational fields, $\lambda$ denotes the polarization states, $e^{(\lambda)}_{ij}(\vec p)$ are polarization tensors, $\omega=c|\vec p|= c(p_{x}^2+p_{y}^2+p_{z}^2)^{1/2}$, $\vec{p}_{-}=\{p_x,p_y,-p_z\}$, and H.c. denotes the Hermitian conjugate.

Assume that the object A is in its ground state, and denote the vacuum state of the gravitational field as $|0\rangle$. Then, the initial state of the system considered is
\beq\label{phi}
|\phi\rangle=|e_{0}\rangle|0\rangle\;.
\eeq
To derive the dynamical quantum gravitational CP interaction potential, we employ the second-order DDC method to study the boundary-dependent energy shift of object A. This formalism enables us to separately identify the contributions of the \emph{vacuum fluctuations} (\emph{vf}) and  \emph{radiation reaction} (\emph{rr}) to the interaction energy. The effective Hamiltonian governing the time evolution of object A's observables is given by the sum of the following two terms~\cite{yongs2020epjc}
\beq
(H^{eff}_{A})_{vf}=-\frac{i}{8\hbar}\int^{t}_{t_{0}}d t' C^{F}_{ijkl}(\vec r_A(t),\vec r_A(t')) [Q^{F}_{ij}(t),Q^{F}_{kl}(t')]\;,
\eeq
\bea
(H^{eff}_{A})_{rr}=-\frac{i}{8\hbar}\int^{t}_{t_{0}}d t' \chi^{F}_{ijkl}(\vec r_{A}(t),\vec r_{A}(t')) \{Q^{F}_{ij}(t),Q^{F}_{kl}(t')\} \;,
\eea
where $Q^{F}_{ij}$ denotes the free part of $Q_{ij}$, which presents even in the absence of interaction.
Here $C^{F}_{ijkl}(\vec r_{A}(t),\vec r_{A}(t'))$ and $\chi^{F}_{ijkl}(\vec r_{A}(t),\vec r_{A}(t'))$ are two statistical functions for the fluctuating gravitational fields in vacuum, which take the form
\beq\label{Cijkl}
C^{F}_{ijkl}(\vec r(t),\vec r(t'))=\frac{1}{2}\langle 0|\{E^{F}_{ij}(\vec r(t)),E^{F}_{kl}(\vec r(t')) \} |0\rangle\;,
\eeq
\beq\label{Fijkl}
\chi^{F}_{ijkl}(\vec r(t),\vec r(t'))=\frac{1}{2}\langle 0|[E^{F}_{ij}(\vec r(t)),E^{F}_{kl}(\vec r(t'))] |0\rangle\;,
\eeq
where $E^{F}_{ij}$ denotes the free part of $E_{ij}$.
To obtain the energy shift of object A, we now evaluate the average values of the effective Hamiltonians $(H^{eff}_{A})_{vf}$ and $(H^{eff}_{A})_{rr}$ on the ground state $|e_0\rangle$, and obtain
\beq\label{Evf}
(\Delta E_{A})_{vf}=-\frac{i}{4\hbar}\int^{t}_{t_{0}}d t' C^{F}_{ijkl}(\vec r_{A}(t),\vec r_{A}(t')) \chi_{A}^{ijkl}(t,t')\;,
\eeq
\beq\label{Esr}
(\Delta E_{A})_{rr}=-\frac{i}{4\hbar}\int^{t}_{t_{0}}d t' \chi^{F}_{ijkl}(\vec r_{A}(t),\vec r_{A}(t')) C_{A}^{ijkl}(t,t')\;,
\eeq
where $\chi_{A}^{ijkl}(t,t')$ and $C_{A}^{ijkl}(t,t')$ are respectively the antisymmetric and symmetric statistical functions of object A, which take the form
\beq
\chi_{A}^{ijkl}(t,t')=\frac{1}{2}\langle e_0|[Q^{F}_{ij}(t),Q^{F}_{kl}(t')]|e_0\rangle\;,
\eeq
\beq
C_{A}^{ijkl}(t,t')=\frac{1}{2}\langle e_0|\{Q^{F}_{ij}(t),Q^{F}_{kl}(t')\}|e_0\rangle\;.
\eeq
It shows that the dynamical gravitational CP interaction between object A and a gravitational Dirichlet boundary arises from the combined effects of the \emph{vacuum fluctuations} and \emph{radiation reaction} (i.e., the radiation field generated by the vacuum-fluctuation-induced quadrupole moment). Thus, the time-dependent interaction energy shift for object A can be expressed as
\beq\label{E_A}
\Delta E_{A}=-\frac{i}{4\hbar}\int^{t}_{t_{0}}d t' \left[C^{F}_{ijkl}(\vec r_{A}(t),\vec r_{A}(t')) \chi_{A}^{ijkl}(t,t')+\chi^{F}_{ijkl}(\vec r_{A}(t),\vec r_{A}(t')) C_{A}^{ijkl}(t,t')\right].
\eeq
Physically, the gravitational CP interaction originates from the absorption or emission of gravitons, which either emerge in pure vacuum or are subsequently reflected by the gravitational boundary. As a result, the interaction energy between a nonpointlike object and a gravitational boundary is determined by both the gravitational vacuum fluctuations and the field radiated by the induced quadrupole moment.

\section{Dynamical gravitational CP interaction}
To obtain the explicit expression of $\Delta E_{A}$, the statistical functions $C^{F}_{ijkl}(\vec r_{A}(t),\vec r_{A}(t'))$ and  $\chi^{F}_{ijkl}(\vec r_{A}(t),\vec r_{A}(t'))$ of the gravitational field are now evaluated. According to Eqs. (\ref{Eij}), (\ref{hij}), (\ref{Cijkl}) and (\ref{Fijkl}), and keep only the boundary-dependent terms, we get
\bea\label{Cijkl1}
\nonumber && C^{F}_{ijkl}(\vec r_A(t),\vec r_A(t'))\\
\nonumber &&=-\frac{\hbar c G}{16\pi^2}\int d^3 \vec p  p^3\sum_{\lambda} \left[\hat e_{ij}^{(\lambda)}(\vec p) \hat e_{kl}^{(\lambda)}(\vec p_{-}) e^{i(\vec p_{-}-\vec p)\cdot \vec r_A} +\hat e_{ij}^{(\lambda)}(\vec p_{-}) \hat e_{kl}^{(\lambda)}(\vec p) e^{i(\vec p-\vec p_{-})\cdot \vec r_A}\right] \left(e^{-i\omega\Delta t'}+ e^{i\omega\Delta t'} \right)  \\
&&=-\frac{\hbar c G}{16\pi^2}\int d^3 \vec p  p^3\sum_{\lambda} \left[\hat e_{ij}^{(\lambda)}(\vec p) \hat e_{kl}^{(\lambda)}(\vec p_{-}) e^{-i\vec p \cdot \vec r}+\hat e_{ij}^{(\lambda)}(\vec p_{-})\hat e_{kl}^{(\lambda)}(\vec p) e^{i\vec p\cdot\vec r} \right]\left(e^{-i\omega\Delta t'}+e^{i\omega\Delta t'} \right) \;,
\eea
and
\bea\label{Fijkl1}
\nonumber && \chi^{F}_{ijkl}(\vec r_A(t),\vec r_A(t'))\\
\nonumber &&=-\frac{\hbar c G}{16\pi^2}\int d^3 \vec p  p^3\sum_{\lambda} \left[\hat e_{ij}^{(\lambda)}(\vec p) \hat e_{kl}^{(\lambda)}(\vec p_{-}) e^{i(\vec p_{-}-\vec p)\cdot \vec r_A} +\hat e_{ij}^{(\lambda)}(\vec p_{-}) \hat e_{kl}^{(\lambda)}(\vec p) e^{i(\vec p-\vec p_{-})\cdot \vec r_A}\right] \left(e^{-i\omega\Delta t'}- e^{i\omega\Delta t'} \right)  \\
&&=-\frac{\hbar c G}{16\pi^2}\int d^3 \vec p  p^3\sum_{\lambda} \left[\hat e_{ij}^{(\lambda)}(\vec p) \hat e_{kl}^{(\lambda)}(\vec p_{-}) e^{-i\vec p \cdot \vec r}+\hat e_{ij}^{(\lambda)}(\vec p_{-})\hat e_{kl}^{(\lambda)}(\vec p) e^{i\vec p\cdot\vec r} \right]\left(e^{-i\omega\Delta t'}-e^{i\omega\Delta t'} \right) \;,
\eea
where $\Delta t'=t-t'$ and $\vec r=(0,0,2 z_A)$. Here the summation of polarization tensors in the transverse traceless gauge gives~\cite{yu1999,yu2018}
\bea
\nonumber\sum_{\lambda}e^{(\lambda)}_{ij}(\vec p) e^{(\lambda)}_{kl}(\vec p')=&&\delta_{ik}\delta_{j l}+\delta_{il}\delta_{j k}-\delta_{ij}\delta_{k l}+\hat p_{i}\hat p_{j}\hat p^{'}_{k}\hat p'_{l}+\hat p_{i}\hat p_{j}\delta_{k l}\\
&&+\hat p'_{k}\hat p'_{l}\delta_{ij}-\hat p_{i}\hat p'_{k}\delta_{j l}-\hat p_{i}\hat p'_{l}\delta_{j k}-\hat p_{j}\hat p'_{k}\delta_{i l}-\hat p_{j}\hat p'_{l}\delta_{i k} \;,
\eea
where $\hat p_i$ is the $i$-th component of the unit vector $\vec p /|\vec p|$. From the summation of polarization tensors, we get
\bea\label{Hr}
\nonumber\sum_{\lambda}e^{(\lambda)}_{ij}(\vec p) e^{(\lambda)}_{kl}(\vec p_{-}) e^{i \vec p \cdot \vec r} &=&\frac{1}{p^4}\sigma_{km}\sigma_{ln}[(\delta_{im}\delta_{j n}+\delta_{i n}\delta_{j m} -\delta_{ij}\delta_{m n})\nabla^4 +(\partial_i\partial_j\delta_{m n} +\partial_m\partial_n\delta_{ij} \\
\nonumber &&-\partial_i\partial_m\delta_{j n}-\partial_i\partial_n\delta_{j m}-\partial_j\partial_m\delta_{i n} -\partial_j\partial_n\delta_{i m}) \nabla^2 +\partial_i\partial_j\partial_m\partial_n]e^{i \vec p \cdot \vec r} \\
&=& \frac{1}{p^4}\sigma_{km}\sigma_{ln} H_{ijmn} e^{i \vec p \cdot \vec r}\;,
\eea
where $H_{ijmn}$ denotes the differential operator, $\sigma_{km}=\delta_{km}-2\delta_{k3}\delta_{m3}$, and $\nabla^2=\partial_i\partial^i$. Substituting Eq. (\ref{Hr}) into Eqs. (\ref{Cijkl1}) and (\ref{Fijkl1}), and performing the integration in the spherical coordinate, we obtain
\bea\label{Cijkl2}
\nonumber C^{F}_{ijkl}(\vec r_A(t),\vec r_A(t'))&=&-\frac{\hbar c G}{2\pi}\sigma_{km}\sigma_{ln}H_{ijmn}  \int_{0}^{\infty}  \frac{\sin{p r}}{r} \left(e^{-i\omega\Delta t'}+e^{i\omega\Delta t'} \right) d p  \\
\nonumber &=&\frac{i\hbar c G}{2}\sigma_{km}\sigma_{ln}H_{ijmn} \frac{1}{2\pi r} \int_{-\infty}^{\infty} \left[e^{i p (r-c\Delta t')}+e^{i p (r+c\Delta t')} \right] d p \\
&=&\frac{i\hbar c G}{2}\sigma_{km}\sigma_{ln}H_{ijmn} \frac{1}{r} [\delta(r-c\Delta t')-\delta(r+c\Delta t')] \;,
\eea
and
\bea\label{Fijkl2}
\nonumber \chi^{F}_{ijkl}(\vec r_A(t),\vec r_A(t'))&=&-\frac{\hbar c G}{2\pi}\sigma_{km}\sigma_{ln}H_{ijmn}  \int_{0}^{\infty}  \frac{\sin{p r}}{r} \left(e^{-i\omega\Delta t'}-e^{i\omega\Delta t'} \right) d p  \\
&=&-\frac{\hbar c G}{2\pi}\sigma_{km}\sigma_{ln}H_{ijmn} \frac{2}{r^2-c^2\Delta t'^2}   \;,
\eea
As for the statistical functions $\chi_{A}^{ijkl}(t,t')$ and $C_{A}^{ijkl}(t,t')$ of the object A, their forms can be obtained as
\bea
\chi_{A}^{ijkl}(t,t')&=&\frac{1}{2}\sum_{s} \hat Q^{s0}_{ij}\hat Q^{0s}_{kl}(e^{-i\omega_{s}\Delta t'} -e^{i\omega_{s}\Delta t'})\;, \label{chiA} \\
C_{A}^{ijkl}(t,t')&=&\frac{1}{2}\sum_{s} \hat Q^{s0}_{ij}\hat Q^{0s}_{kl}(e^{-i\omega_{s}\Delta t'} +e^{i\omega_{s}\Delta t'}) \label{CA} \;,
\eea
where $\hat Q^{s0}_{ij}=e^{i\omega_{s}t}\langle e_{s}|Q^{F}_{ij}|e_{0}\rangle$, $\hat Q^{0s}_{ij}=e^{-i\omega_{s}t}\langle e_{0}|Q^{F}_{ij}|e_{s} \rangle$, and $\hat Q^{s0}_{ij}=\hat Q^{0s}_{ij}$ is assumed.

\begin{figure}[htbp]
  \centering
  \includegraphics[width=0.5\textwidth]{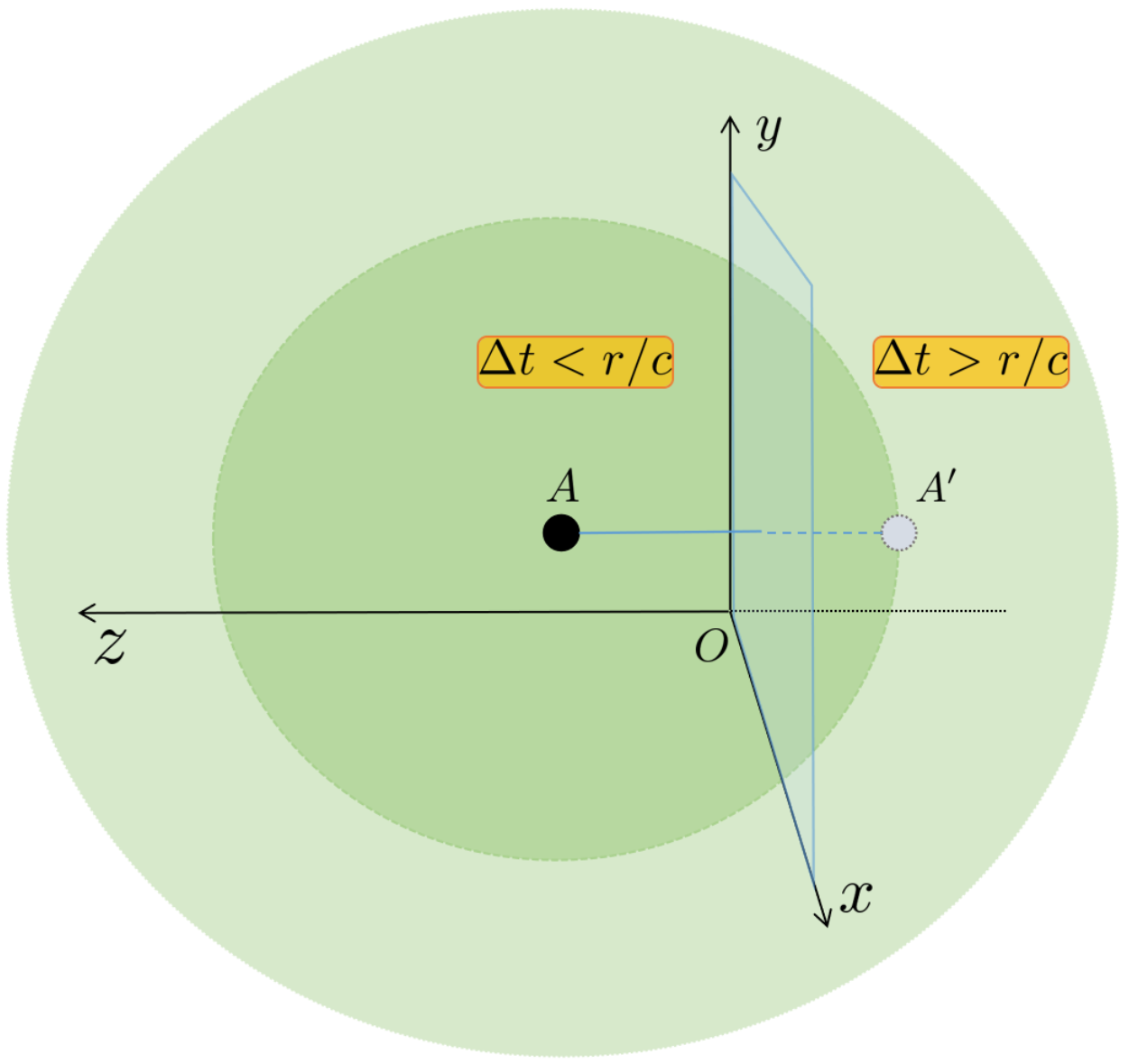}\\
  \caption{A schematic diagram for the dynamical quantum gravitational CP interaction between object $A$ and its mirror image $A'$. The regions $\Delta t<r/c$ and $\Delta t>r/c$ respectively corresponds to the two different colored areas. And the gravitational plane boundary is located at $z=0$ and acts as the mirror.}\label{P1}
\end{figure}

Substituting Eqs. (\ref{Cijkl2}), (\ref{Fijkl2}), (\ref{chiA}) and (\ref{CA}) into Eq. (\ref{E_A}), and performing the integration of $t'$, the time-dependent gravitational CP interaction is thus obtained
\bea\label{E_A dyn}
\nonumber \Delta E_A &=& \frac{\hbar c G}{8\pi\hbar c} \sum_{s}\hat Q^{s0}_{ij}\hat Q^{0s}_{kl} \sigma_{km}\sigma_{ln} H_{ijmn} \frac{1}{r}\Big\{[2\text{Ci}(p_s r)-\text{Ci}(p_s r-c p_s\Delta t)\theta{(r-c\Delta t)}\\
\nonumber &&-\text{Ci}(c p_s\Delta t-p_s r)\theta{(c\Delta t -r)} -\text{Ci}(p_s r+c p_s\Delta t)]\sin{p_s r} -[2\text{Si}(p_s r)  \\
&&-\text{Si}(p_s r-c p_s\Delta t) - \text{Si}(p_s r+c p_s\Delta t) - \pi\theta{(c\Delta t -r)}]\cos{p_s r}  \Big\}\;,
\eea
where $\Delta t=t-t_0$ and $p_s=\omega_s/c$. Here $\text{Ci}(x)$ and $\text{Si}(x)$ are cosine and sine integrals, $\theta(x)$ is a step function, which equals to $1$ when $x>0$ but $0$ when $x<0$.
Eq. (\ref{E_A dyn}) shows that the quantum gravitational CP interaction between object A and a gravitational Dirichlet boundary, or in other words, between the object and its mirror image (with the plane boundary acting as the mirror and $r$ denoting the distance), exhibits time dependence and is therefore referred to as the dynamical gravitational CP  interaction. Furthermore, the step function $\theta(x)$ separate the formalism of $\Delta E_A$ into two kinds, corresponding to two temporal regions: $\Delta t<r/c$ and $\Delta t>r/c$, i.e., the mirror image of the object is respectively located outside and inside its light cone. A schematic diagram is shown in Fig.~\ref{P1}.

In the following, let us discuss some physical characteristics of the dynamical gravitational CP interaction.
First, when $\Delta t<r/c$, the radiation generated by the gravitational vacuum-fluctuation-induced mass quadrupole of the object has not yet been reflected by the gravitational boundary and back-reacted to the object, i.e., the mirror image of the object lies outside its light cone. The interaction potential Eq. (\ref{E_A dyn}) becomes
\bea \label{E_A out}
\nonumber \Delta E_A &=& \frac{\hbar c G}{8\pi\hbar c} \sum_{s}\hat Q^{s0}_{ij}\hat Q^{0s}_{kl} \sigma_{km}\sigma_{ln} H_{ijmn} \frac{1}{r}\Big\{[2\text{Ci}(p_s r)-\text{Ci}(p_s r-c p_s\Delta t)-\text{Ci}(p_s r+c p_s\Delta t)]\\
&& \times \sin{p_s r}-[2\text{Si}(p_s r)-\text{Si}(p_s r-c p_s\Delta t) - \text{Si}(p_s r+c p_s\Delta t)]\cos{p_s r}  \Big\}\;.
\eea
Eq. (\ref{E_A out}) shows that a nonzero interaction potential exists between an object and its mirror image, even when the image is located outside the causal region of the object. Obviously, the current result is significantly different to the interobject potential between two entangled nonpointlike object placed near a gravitational boundary, which appears only when one object (or its image) lies within the light cone of the other \cite{yongs2023}.
Such an apparently unexpected result actually reflects the nonlocality of the fluctuating gravitational field in vacuum. This is because, within the temporal region where $\Delta t<r/c$, the contribution to the dynamical gravitational CP interaction potential arises only from the vacuum fluctuations described in Eq. (\ref{Evf}).
Moreover, the dynamical interaction potential Eq. (\ref{E_A out}) can be positive (repulsive) or negative (attractive), depending on the interaction time $\Delta t$ and the object-image distance $r$, in contrast to the static form Eq. (\ref{E_A sta}) which is always positive in both the near and far regimes.

Second, when $\Delta t>r/c$, the radiation generated by the gravitational vacuum-fluctuation-induced mass quadrupole of the object has been reflected by the gravitational boundary and back-reacted to the object, i.e., the mirror image of the object is located within its light cone. The interaction potential Eq. (\ref{E_A dyn}) becomes
\bea\label{E_A in}
\nonumber \Delta E_A &=& \frac{\hbar c G}{8\pi\hbar c} \sum_{s}\hat Q^{s0}_{ij}\hat Q^{0s}_{kl} \sigma_{km}\sigma_{ln} H_{ijmn} \frac{1}{r}\Big\{[2\text{Ci}(p_s r)-\text{Ci}(c p_s\Delta t-p_s r)-\text{Ci}(p_s r+c p_s\Delta t)]\\
&& \times \sin{p_s r} -[2\text{Si}(p_s r)-\text{Si}(p_s r-c p_s\Delta t) - \text{Si}(p_s r+c p_s\Delta t)-\pi]\cos{p_s r}  \Big\}\;.
\eea
It shows that the dynamical gravitational CP interaction potential remains time-dependent and correlates with the object-image distance $r$, when the mirror image of the object lies within its causal region.
Specifically, in the temporal region $\Delta t>r/c$, the interaction potential receives contributions from both vacuum fluctuations and radiation reaction, whereas the time-dependence arises solely from the vacuum fluctuations.
In the limit of large times, specifically as $\Delta t\rightarrow\infty$, Eq. (\ref{E_A in}) asymptotically approaches
\beq \label{E_A sta}
\Delta E_A = \frac{\hbar c G}{4\pi\hbar c}\sum_{s}\hat Q^{s0}_{ij}\hat Q^{0s}_{kl} \sigma_{km}\sigma_{ln} H_{ijmn} \frac{1}{r}\left\{\text{Ci}(p_s r)\sin{p_s r} -\left[\text{Si}(p_s r)-\frac{\pi}{2}\right]\cos{p_s r}  \right\} \;,
\eeq
which is actually the static gravitational CP interaction potential.
In the near regime, i.e., $p_s r \ll 1$, the leading term of the potential Eq. (\ref{E_A sta}) takes the form
\beq
\Delta E_A = \frac{\hbar c G}{8\hbar c}\sum_{s}\hat Q^{s0}_{ij}\hat Q^{0s}_{kl} \sigma_{km}\sigma_{ln} H_{ijmn} \frac{1}{r} \;,
\eeq
Utilize the integral representation
$
\pi=\int_{0}^{\infty}\frac{2p_s}{p_s^2+u^2} d u,
$
and introduce the gravitoelectric polarizability of the nonpointlike object
$
\alpha_{ijkl}(\omega)=\sum_{s} \frac{2\hbar\omega_s \hat Q^{s0}_{ij} \hat Q^{0s}_{kl}}{(\hbar\omega_s)^2-(\hbar \omega)^2}\;,
$
we obtain
\beq \label{E_A sta n}
\Delta E_A = \frac{\hbar c G}{8\pi}\sigma_{km}\sigma_{ln} H_{ijmn} \frac{1}{r}\int_{0}^{\infty} \alpha_{ijkl}(i u) du \;.
\eeq
For simplicity,  we assume that the object A is isotropically polarizable, i.e.,
$
\alpha_{ijkl}(\omega)=(\delta_{ik}\delta_{jl}+\delta_{il}\delta_{jk})\alpha(\omega)\;,
$
where $\alpha(\omega)$ denotes the isotropic polarizability. Then, Eq. (\ref{E_A sta n}) can be simplified as
\beq
\Delta E_A = \frac{1581\hbar c G}{2^7\pi z_A^5}\int_{0}^{\infty} \alpha(i u) du \;,
\eeq
where $r=2z_A$ has been used.
In the far regime, i.e., $p_s r \gg 1$, the leading term of the potential Eq. (\ref{E_A sta}) takes the form
\bea \label{E_A sta f}
\Delta E_A &=& \frac{\hbar c G}{4\pi\hbar c}\sum_{s}\hat Q^{s0}_{ij}\hat Q^{0s}_{kl} \sigma_{km}\sigma_{ln} H_{ijmn} \frac{1}{p_s r^2}
= \frac{195\hbar c G}{8\pi z_A^6}\alpha(0)\;.
\eea
The above results demonstrate that the dynamical gravitational CP interaction reduces to the static one when the interaction time is sufficiently long, allowing the system to approach asymptotic equilibrium. It is worth noting that the static result (\ref{E_A sta}) remains positive in both the near and far regimes, signifying a repulsive force between the object and the gravitational boundary. Moreover, the asymptotic behaviors of the static gravitational CP interaction in both the near and far regimes agree with that reported in Ref. \cite{Hu2017}, although the coefficient of the interaction potential is different. This discrepancy arises from the distinct treatment of the gravitoelectric polarizability and the summation of polarization tensors. In the present study, we adopt the method used in Ref. \cite{yu2018} to calculate the summation of polarization tensors, which explicitly incorporates the inversion of polarization. Besides, a general definition of the gravitoelectric polarizability $\alpha_{ijkl}(\omega)$ for a nonpointlike object is employed.

\section{Discussion}
\label{sec_disc}
In this paper, the dynamical gravitational CP interaction between a nonpointlike object and a plane gravitational Dirichlet boundary is explored in the framework of linearized quantum gravity, based on the second-order DDC formalism. The analytical expression for the time-dependent quantum gravitational CP interaction potential is derived, and we find that the dynamical potential remains nonzero regardless of whether the radiation generated by the gravitational vacuum-fluctuation-induced mass quadrupole of the object is reflected by the gravitational boundary and back-reacted to the object. In other words, the dynamical gravitational CP interaction occurs irrespective of whether the mirror image of the object lies outside or inside its causal region. This result implies the nonlocality of the fluctuating gravitational field in vacuum and calls for a reevaluation of the intrinsic causality within this interaction.
Moreover, the dynamical gravitational CP interaction can be either attractive or repulsive depending on the distance of the object with respect to the boundary and the duration of the interaction.
When the interaction time is sufficiently long for the system to approach asymptotic equilibrium, the dynamical potential reduces to the static one, which is time-independent and is consistently repulsive in both the near and far regimes.

\begin{acknowledgments}

This work was supported in part by the NSFC under Grants No.12405051 and No. 12205032, by the Chongqing Human Resources and Social Security Administration Program under Grants No. D63012024013 and No. cx2021044, and by the Talent Introduction Program of Chongqing University of Posts and Telecommunications under Grants No. E012A2024028 and No. E012A2020248.

\end{acknowledgments}

\end{document}